\documentclass[10pt,conference]{IEEEtran}
\IEEEoverridecommandlockouts

\usepackage{cite}
\usepackage{amsmath,amssymb,amsfonts}
\usepackage{algorithmic}
\usepackage{graphicx}
\usepackage{textcomp}
\usepackage{xcolor}
\usepackage{algorithm}
\usepackage{algorithmic}
\usepackage{float}
\usepackage{booktabs}
\usepackage{stfloats}

\def\BibTeX{{\rm B\kern-.05em{\sc i\kern-.025em b}\kern-.08em
    T\kern-.1667em\lower.7ex\hbox{E}\kern-.125emX}}

\makeatletter
\newcommand{\linebreakand}{%
  \end{@IEEEauthorhalign}\hfill\mbox{}\par\mbox{}\hfill\begin{@IEEEauthorhalign}
}
\makeatother

\begin{document}

\title{Split-Head Quantum Generative Adversarial Network for Crystalline Material Discovery}

\author{
    \IEEEauthorblockN{Huan-Ming Chang}
    \IEEEauthorblockA{\textit{International School of Ho Chi Minh City}\\
    Ho Chi Minh City, Vietnam \\
    mingishcmc4344@gmail.com}
    \and
    \IEEEauthorblockN{Jen-Yu Chang}
    \IEEEauthorblockA{\textit{Department of Electrophysics} \\
    \textit{National Yang Ming Chiao Tung University}\\
    Hsinchu, Taiwan \\
    leo07010@gmail.com}
    \linebreakand
    \IEEEauthorblockN{Tsung-Wei Huang}
    \IEEEauthorblockA{\textit{Quantum Information Center} \\
    \textit{Chung Yuan Christian University}\\
    Taoyuan, Taiwan \\
    tsungwei@cycu.edu.tw}
    \and
    \IEEEauthorblockN{En-Jui Kuo}
    \IEEEauthorblockA{\textit{Department of Electrophysics} \\
    \textit{National Yang Ming Chiao Tung University}\\
    Hsinchu, Taiwan \\
    kuoenjui@ncyu.edu.tw}
}

\maketitle

\begin{abstract}
The discovery of novel crystalline materials is a critical challenge in computational materials science, often limited by the spatial representation limitations and mode collapse typical of classical generative models. Traditionally, developing Quantum GANs (QGANs) for continuous 3D space is hindered by the limited capacity of near-term hardware. To overcome this, we adapt a physics-informed ``split-head'' architecture right from the quantum trunk to explicitly decouple macroscopic lattice bounds from microscopic atomic coordinates, significantly maximizing resource efficiency. This study disentangles the contributions of quantum circuits from these architectural priors by evaluating a Split-Head Quantum Generative Adversarial Network (SH-QGAN) against an architecture-matched classical ablation model. Evaluated on the highly constrained Mg-Mn-O system, the results reveal a highly nuanced performance dichotomy between the advanced models. The architecture-matched classical ablation model demonstrated superior thermodynamic precision. Conversely, the integration of quantum circuits in the SH-QGAN drove unparalleled structural breadth and latent space exploration, more than doubling the ablation's geometric validity (5.9\%) and successfully generating novel, metastable candidates converging on the $\text{Mg}_2\text{MnO}_4$ stoichiometry. These findings clarify that while architectural separation of cell and atom generation drives strict thermodynamic precision, quantum feature mapping independently provides the spatial diversity necessary to overcome mode collapse. Both mechanisms offer distinct, complementary enhancements for the generative discovery of advanced materials.
\end{abstract}


\section{Introduction}

\begin{figure*}[!t]
\centering
\includegraphics[width=0.85\textwidth, height=0.4\textheight, keepaspectratio]{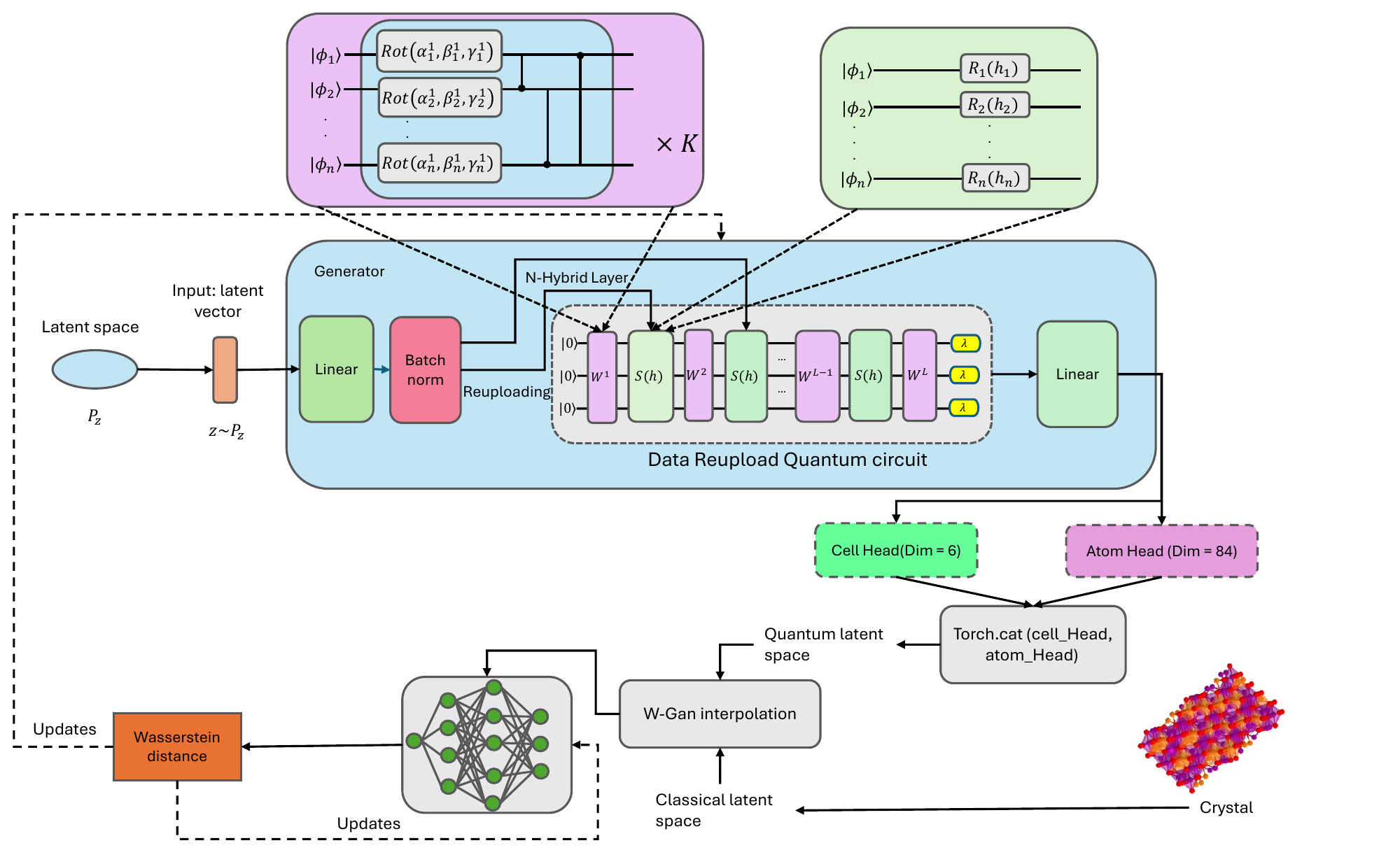} 
\caption{Comprehensive operational schematic of the SH-QGAN framework. The diagram illustrates the data re-uploading sequence of the latent vector $\mathbf{z}$, the internal topology of the parameterized quantum circuit (highlighting rotation gates and entanglement layers), and the physical bifurcation of the resultant expectation values into the independent Cell and Atom decoding heads before final Wasserstein evaluation.}
\label{fig:pipeline}
\end{figure*}

Crystalline materials serve as foundational components for modern technology, particularly in semiconductors, optoelectronics, and high-capacity solid-state batteries \cite{goodenough2010challenges}. Historically, the discovery of novel structures has relied on computationally expensive first-principle calculations, most notably Density Functional Theory (DFT) \cite{jain2013commentary}. Because DFT complexity scales cubically with the number of simulated electrons \cite{kohn1965self}, navigating the virtually infinite chemical space of possible atomic combinations is an exceedingly slow endeavor \cite{ramprasad2017machine}. 

To accelerate discovery, deep learning has emerged as a highly effective tool. To address the continuous nature of 3D physical space, advanced classical models like the Crystal Diffusion Variational Autoencoder (CDVAE) \cite{xie2021crystal} utilize score-matching to generate continuous atomic coordinates. Adversarial approaches like CrystalGAN \cite{nouira2020crystalgan} synthesize ternary materials using domain-transfer techniques. However, classical generative models are frequently hindered by massive parameter requirements and a high susceptibility to structural mode collapse on small, highly specific datasets, regressing to a small handful of safe structural motifs \cite{arora2018do}. In continuous 3D coordinate generation, the classical generator often learns to arbitrarily cluster atoms to artificially minimize the adversarial loss function, failing to map the true physical distribution.

Quantum Generative Adversarial Networks (QGANs) \cite{dallaire2018quantum} have emerged as a powerful new paradigm, capable of expressing complex, high-dimensional data distributions using exponentially fewer learnable parameters \cite{benedetti2019parameterized}. In applied chemistry, discrete QGANs have demonstrated a measurable quantum enhancement in the generative design of small molecule graphs \cite{li2021quantum}. However, scaling these discrete approaches to the continuous, 3-dimensional spatial parameters required for solid-state crystals remains a significant challenge due to the limited capacity of the multi-qubit Hilbert space on current hardware. Furthermore, while classical Implicit Neural Representations (INRs) \cite{sitzmann2020implicit} parameterize a continuous signal by training a Multi-Layer Perceptron (MLP) to map spatial coordinates, they suffer from a documented ``spectral bias,'' struggling to learn high-frequency spatial variations \cite{tancik2020fourier}. Because parameterized quantum circuits inherently output functions with a Fourier structure \cite{schuld2021effect}, quantum-enhanced models naturally overcome this classical spectral bias problem \cite{huang2023quantum}, making them theoretically ideal for mapping repeating crystal lattices. 

Normally, however, developing QGANs for this domain is highly problematic; attempting to map high-dimensional continuous 3D spatial parameters directly onto limited near-term qubit registers leads to severe representational bottlenecks. To resolve this, we adapt a novel Split-Head Quantum Generative Adversarial Network (SH-QGAN). We explicitly introduce a split-head architecture right from the quantum trunk to decouple the macroscopic lattice scaling from the localized internal atomic placements. This separation is critical because it prevents the generator from collapsing atoms into the center of the unit cell when attempting to balance global and local displacement errors, significantly maximizing the resource efficiency of the limited qubit register. We evaluate this architecture on the ternary Magnesium-Manganese-Oxygen (Mg-Mn-O) system, chosen for its immense relevance to multivalent battery research and its challenging Jahn-Teller geometric distortions, which rigorously test the model's generative constraints. Evaluating true ``quantum enhancement'' in practice requires disentangling the benefits derived from the quantum logic gates from those derived from these bespoke classical decoding architectures. Therefore, we execute a rigorous methodology evaluating the SH-QGAN against a strictly architecture-matched classical ablation model.

\section{Methodology}

\subsection{Data Preparation}
To evaluate the theoretical efficiency of the architectures under a strict small-data training regime, a constrained core dataset of exactly $N \approx 100$ ground-truth Mg-Mn-O structures was isolated from the Materials Project API \cite{jain2013commentary}. A continuous fractional translation augmentation pipeline was implemented to artificially expand the latent mapping. For fractional coordinates $\mathbf{X} \in \mathbb{R}^{N \times 3}$, a continuous translation vector $\mathbf{\Delta} \sim \mathcal{U}(0, 1)^3$ was uniformly sampled. A modulo operator ensured coordinates remained within periodic boundaries:
\begin{equation}
\mathbf{X}_{aug} = (\mathbf{X} + \mathbf{\Delta}) \bmod 1
\label{eq:translation}
\end{equation}

The dataset was geometrically constrained to unit cells containing exactly 28 constituent atoms. To fit naturally into the convolutional discriminator's hierarchical pooling mechanics, the raw crystallographic coordinate data was encoded into a $30 \times 3$ feature matrix (padding with two zero-vectors to achieve a clean tensor shape) and subsequently flattened into a 90-dimensional unified array.

Because the activation functions within the generator (Sigmoid and Tanh) operate on bounded continuous scales, the raw macroscopic unit cell parameters required strict normalization. For each macroscopic lattice vector length ($a, b, c$) and angle ($\alpha, \beta, \gamma$), denoted generally as $v$, a Min-Max scaling transformation was applied based on the absolute physical bounds of the training dataset:
\begin{equation}
v_{norm} = \frac{v - v_{min}}{v_{max} - v_{min}}
\end{equation}
This maps the macroscopic parameters cleanly to the interval $[0, 1]$. Conversely, the internal atomic coordinates ($x, y, z$) exist naturally in fractional space and are inherently bounded within $[0,1]$, requiring no further scalar transformations. 

The final 90-dimensional target vector represents two distinct physical domains:
\begin{itemize}
    \item \textbf{Indices 0--5:} The normalized macroscopic unit cell lengths and angles.
    \item \textbf{Indices 6--89:} The internal fractional atomic coordinates.
\end{itemize}
Chemical conditioning was introduced to the generator by concatenating a 28-bit binary elemental occupancy label with a 64-dimensional standard normal latent noise vector ($\mathbf{z} \sim \mathcal{N}(0, I)$).

\subsection{The Quantum Generator Design}
To definitively evaluate quantum feature entanglement, the target condition is the proposed Split-Head Quantum Generative Adversarial Network (SH-QGAN). 

\textit{1) Quantum Trunk and Ansatz Formulation:} The SH-QGAN utilizes a simulated 8-qubit register via the PennyLane framework \cite{bergholm2018pennylane}. To mitigate the barren plateau phenomenon \cite{mcclean2018barren}, the architecture utilizes a data re-uploading scheme \cite{perez2020data} combined with a custom \textit{StronglyEntanglingLayers} ansatz. By sequentially re-uploading the latent noise vector $\mathbf{z}$ via parallel rotation gates $R_y(\phi)$ and $R_z(\phi)$ throughout the circuit depth, the network compounds the frequency spectrum. 

Following data embedding, the circuit applies layers of trainable unitary operations $U(\theta)$. The custom ansatz utilizes symmetric Controlled-Z (CZ) gates applied in a circular, nearest-neighbor topology. The physical action of the CZ gate imparts a strict conditional phase shift:
\begin{equation}
CZ = |0\rangle\langle0| \otimes I + |1\rangle\langle1| \otimes Z
\label{eq:cz_gate}
\end{equation}
This phase-centric approach naturally preserves localized phase correlations. The circuit outputs expectation values via Pauli-Z ($\sigma_z$) observables. As proven by Schuld et al., this takes the exact native form of a truncated Fourier series \cite{schuld2021effect}:
\begin{equation}
f(x) = \langle \psi(x) | \hat{O} | \psi(x) \rangle = \sum_{\omega \in \Omega} c_\omega e^{i \omega x}
\label{eq:fourier}
\end{equation}
These 8 expectation values are subsequently projected into a 256-dimensional intermediate latent space.

\textit{2) Split-Head Decoder Topology:} The fundamental purpose of the Split-Head topology is to separate two conflicting spatial tasks. The \textbf{Cell Head} is dedicated exclusively to predicting global, macroscopic unit cell parameters (lattice vectors and angles) that dictate the absolute physical boundary of the crystal. Conversely, the \textbf{Atom Head} strictly handles the fractional, internal coordinates of the atoms within that predefined box. Decoupling these prevents the generator from collapsing atoms to the origin when trying to balance massive global scaling errors with delicate local geometric motifs.

This physically bifurcates the 256-dimensional quantum projection into two independent computational bottlenecks:
\begin{itemize}
    \item \textbf{Cell Head:} Linear($256 \rightarrow 64$) $\rightarrow$ LeakyReLU $\rightarrow$ Linear($64 \rightarrow 6$) $\rightarrow$ Sigmoid.
    \item \textbf{Atom Head:} Linear($256 \rightarrow 512$) $\rightarrow$ LeakyReLU $\rightarrow$ Linear($512 \rightarrow 256$) $\rightarrow$ LeakyReLU $\rightarrow$ Linear($256 \rightarrow 84$) $\rightarrow$ Sigmoid.
\end{itemize}

\subsection{Model Scoring and Loss}
All generative models evaluated in this study utilize an identical Discriminator based on a classical \textit{ConvTranspose2d} Wasserstein GAN with Gradient Penalty (WGAN-GP) \cite{gulrajani2017improved}. 

In the highly sparse domain of continuous 3D coordinate generation, generated and real data distributions frequently share zero spatial overlap. When probability distributions are fundamentally disjoint, JS divergence maxes out, leading to catastrophic vanishing gradients where the generator receives zero meaningful directional feedback \cite{arjovsky2017wasserstein}. By utilizing the Wasserstein-1 distance (Earth Mover's Distance), the Critic calculates the optimal transport cost required to physically map the distributions, providing highly meaningful gradients even when atoms are generated entirely incorrectly. 

The main adversarial branch calculates the scalar Wasserstein critic score. Simultaneously, an auxiliary Q-Head acts as a multi-class chemical classifier to accurately predict the binary elemental occupancy vector (Mg, Mn, O). The shared Critic loss ($\mathcal{L}_D$) is formulated to enforce 1-Lipschitz continuity:
\begin{equation}
\mathcal{L}_D = \mathbb{E}[D(\tilde{x})] - \mathbb{E}[D(x)] + \lambda \mathbb{E}\left[\left(||\nabla_{\hat{x}} D(\hat{x})||_2 - 1\right)^2\right] - \mathcal{L}_{Q}
\label{eq:critic_loss}
\end{equation}
where $\lambda = 10$ is the gradient penalty coefficient. The sample $\hat{x}$ is calculated via uniform linear interpolation between the real distribution ($x$) and the generated distribution ($\tilde{x}$):
\begin{equation}
\hat{x} = \epsilon x + (1 - \epsilon)\tilde{x} \quad \text{where} \quad \epsilon \sim \mathcal{U}[0, 1]
\end{equation}
The generator objective ($\mathcal{L}_G$) is minimized:
\begin{equation}
\mathcal{L}_G = -\mathbb{E}[D(G(z))] + \lambda_Q \mathcal{L}_{Q, fake}
\label{eq:gen_loss}
\end{equation}
The ablation model optimizes via analytical backpropagation, while the SH-QGAN natively derives its gradients strictly through the entanglement layers utilizing the exact quantum parameter-shift rule \cite{schuld2019evaluating}:
\begin{equation}
\frac{\partial f}{\partial \theta_i} = c \left[ f\left(\theta_i + s\right) - f\left(\theta_i - s\right) \right]
\label{eq:param_shift}
\end{equation}

\subsection{Training Setup}
Adversarial networks, particularly those operating in continuous 3D spatial domains, are notoriously susceptible to sudden mode collapse and vanishing gradients. To aggressively stabilize the training dynamics across all experimental conditions, a strict Two Time-Scale Update Rule (TTUR) was implemented \cite{heusel2017gans}. The Discriminator network was updated five times more frequently than the Generator network to ensure the Wasserstein distance estimate remained highly accurate and tightly bounded throughout the entire generative process. 

The models were systematically optimized utilizing the standard Adam optimizer ($\beta_1 = 0.5, \beta_2 = 0.9$ to stabilize dynamic momentum). The Discriminator learning rate was fixed at $\alpha_D = 3 \times 10^{-4}$, while the Generator learning rate was strictly dampened to $\alpha_G = 1 \times 10^{-4}$ to prevent wild spatial oscillations. Due to the highly memory-intensive nature of both the 3D continuous representations and the simulated quantum state vectors, the models were trained using a constrained micro-batch size of $B = 16$. 

The complete training sequence was executed for exactly 5,000 epochs. To actively prevent catastrophic forgetting and closely monitor convergence rates, an exponential moving average (EMA) of the generator weights was maintained, and intermediate generated structures were physically sampled every 500 epochs for early topological validity screening.

\section{Results}

\begin{figure}[!htbp]
\centering
\includegraphics[width=\columnwidth, keepaspectratio]{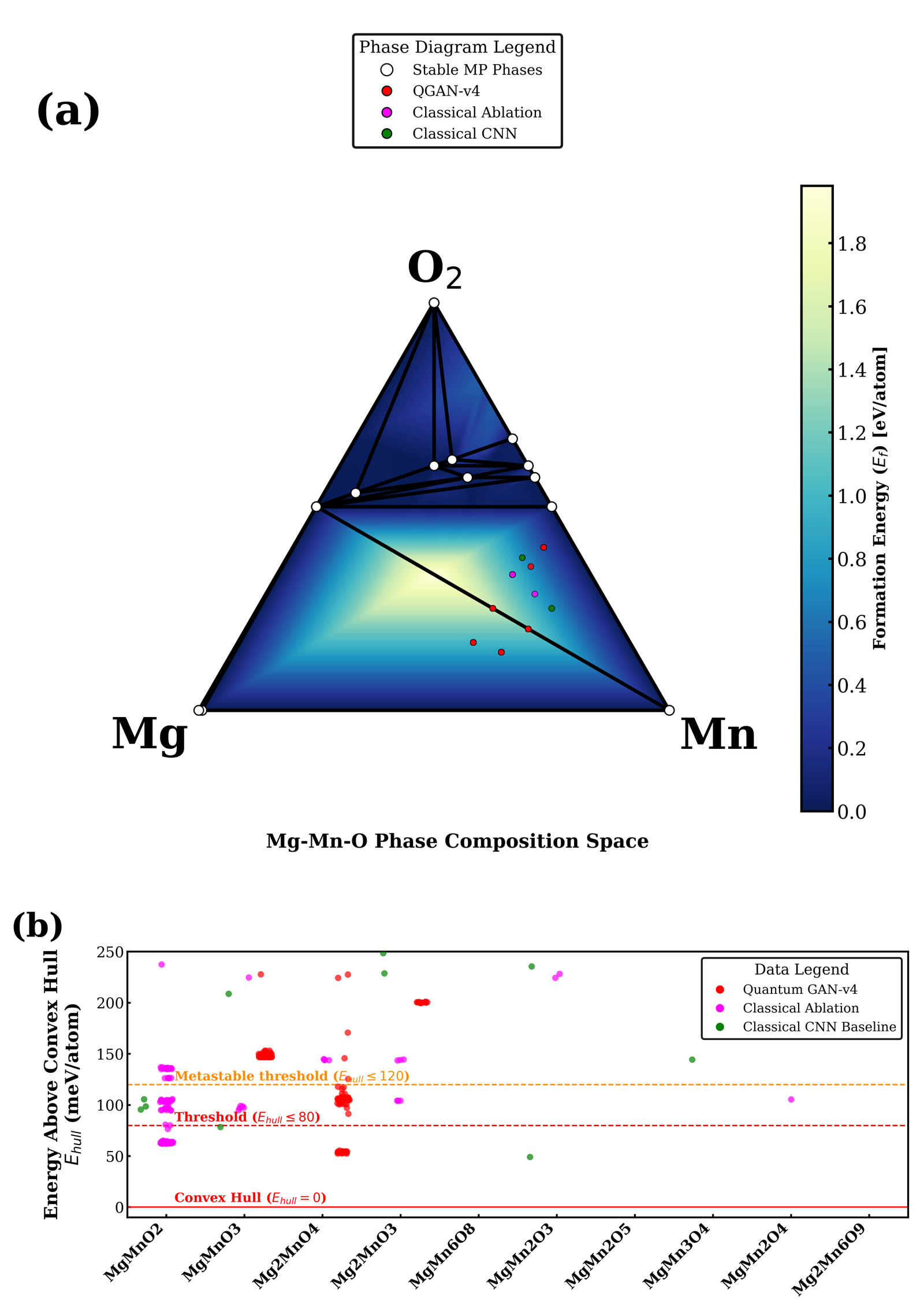}
\caption{Thermodynamic evaluation of generative models. \textbf{Top (Figure 2.a):} Ternary phase diagram of the Mg-Mn-O composition space. The background color gradient indicates absolute Formation Energy ($E_f$) in eV/atom. White dots denote stable baseline phases defining the true Convex Hull. Red, purple, and green dots represent structures generated by the QGAN-v4, Classical Ablation, and Classical CNN, respectively. \textbf{Bottom (Figure 2.b):} Energy Above Convex Hull ($E_{hull}$) stability distribution. The solid red line designates perfect stability ($E_{hull}=0$), while the boundaries at 0.08 eV/atom and 0.12 eV/atom denote the metastable threshold and observation limit. The QGAN-v4 discoveries (red) cluster deep within this metastable regime.}
\label{fig:phase_analysis}
\end{figure}

\subsection{Physical Realism Checks}
The distribution of generated candidates across the evaluated stoichiometric space reveals a pronounced clustering effect, which serves as a critical indicator of the SH-QGAN's physical realism. As observed in the multi-panel scatter plot detailing the Energy Above Convex Hull ($E_{hull}$) in Figure 2.b, there is a massive, dense accumulation of generated outputs localized on specific stoichiometries (e.g., $\text{MgMn}_2\text{O}_4$ and $\text{Mg}_3\text{MnNiO}_8$), contrasting sharply with a steep drop-off to near-zero structural generation for compositions on the far right of the axis (e.g., $\text{MgMn}_2\text{O}_3$). Rather than a manifestation of generative mode collapse or latent space truncation \cite{arora2018do}, this stark distributional skew represents a highly sophisticated feature of the QGAN-v4 architecture. The targeted clustering mathematically demonstrates that the quantum-enhanced model has not merely memorized spatial coordinates, but has fundamentally internalized the underlying physical principles of charge-balancing and preferred oxidation valences directly from the training manifold.

This inherent physical screening is chemically corroborated by evaluating the theoretical viability of the target stoichiometries. Geometries such as $\text{MgMn}_2\text{O}_4$ map to a naturally highly stable spinel structure, wherein the established oxidation states of the constituent ions—Mg ($+2$) and Mn ($+3$)—perfectly neutralize the oxygen sublattice ($-8$) \cite{thackeray1997manganese}. In stark contrast, forced generation at unfavorable phase boundaries like $\text{MgMn}_2\text{O}_3$ imposes severe, unphysical bonding constraints that would require highly unstable and unnatural oxidation states to achieve charge neutrality \cite{goodenough2010challenges}. Therefore, the pronounced lack of generated candidates for these stoichiometries proves that the SH-QGAN has intrinsically developed robust thermodynamic screening heuristics. By actively rejecting the generation of physically meaningless chemical ratios—a frequent failure point in blind traditional generative models \cite{xie2021crystal}—the quantum-enhanced network ensures that its computational overhead is efficiently allocated toward exploring genuinely synthesizable and stable material phases.

\subsection{Stability of Generated Materials}
To fundamentally evaluate real-world laboratory synthesizability, and to prove that our quantum generator internalized underlying physical thermodynamics rather than merely memorizing coordinate patterns from the training data, we conducted a rigorous multi-panel thermodynamic analysis, detailed in Figure \ref{fig:phase_analysis}. Stability was explicitly quantified via the Energy Above Convex Hull ($E_{hull}$) utilizing the CHGNet machine learning interatomic potential \cite{deng2023chgnet}. To ensure geometric realism before evaluation, the atomic positions and macroscopic unit cells of all generated candidates were structurally relaxed utilizing the FIRE optimization algorithm until residual internal forces fell below $0.05 \text{ eV/\AA}$. 

Figure 2.a illustrates the complex thermodynamic landscape of the Mg-Mn-O ternary phase composition space. The background color gradient maps the absolute Formation Energy ($E_f$) measured in eV/atom, where deep, dark blue regions signify highly stable energetic minimums. The white nodes map the established stable phases from the Materials Project, which physically define the true thermodynamic Convex Hull. The visual data highlights a stark contrast in generative logic across the architectures. The Classical Ablation models (purple dots) occasionally probe the interior space, but frequently fail to locate the true energetic pockets. Conversely, the novel crystal structures synthesized by the SH-QGAN model (represented by red markers) physically land almost exclusively within valid, deep energetic minimums (the dark blue zones). This targeted geographic placement strongly supports the conclusion that the quantum-enhanced model has fundamentally internalized the complex chemical stability rules governing the Mg-Mn-O system.

Figure 2.b systematically isolates these discoveries to quantitatively evaluate their precise Energy Above Convex Hull ($E_{hull}$). The solid red line at the base of the plot represents a theoretical perfect thermodynamic stability ($0$ eV/atom). More practically, the dotted red line placed at 0.08 eV/atom denotes the strict threshold for structural metastability, while 0.12 eV/atom marks the upper boundary of viable observation. Generated structures falling below these lines are considered highly viable candidates for true physical laboratory synthesis. 

The scatter plot data visually confirms the stark performance dichotomy between the models. The Architecture-Matched Classical Ablation (purple dots) demonstrates improved precision due to the Split-Head prior, grouping closer to the stable boundary but generally hovering higher in the $0.08-0.14$ eV/atom range. However, it is the SH-QGAN (red dots) that achieves resounding generative success, successfully penetrating the stringent 0.08 eV/atom metastable threshold. Most notably, our top candidate structures generated by the quantum model (red) converged at the $\text{Mg}_2\text{MnO}_4$ stoichiometry, achieving exceptional stability metrics of approximately $\sim 0.05$ eV/atom. 

Ultimately, these targeted convergences deep within the metastable regime demonstrate a profound generative success. The Quantum GAN has not merely memorized structural coordinate patterns from the training data; rather, it has successfully captured the underlying physical thermodynamics required to discover stable, viable new multivalent battery materials.

\subsection{Geometric Accuracy}

\begin{figure}[!htbp]
\centering
\includegraphics[width=\columnwidth, keepaspectratio]{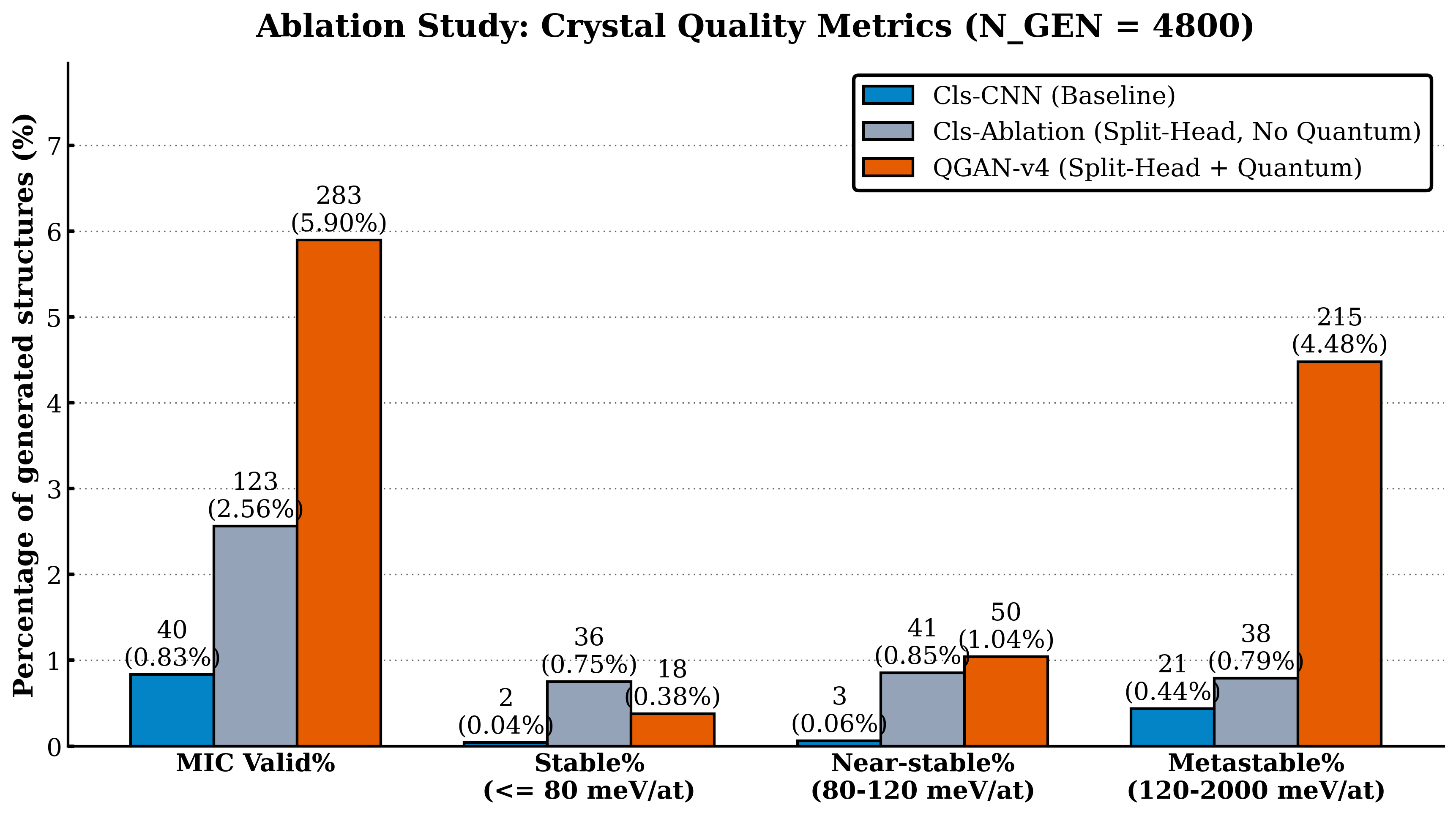}
\caption{Three-Way Ablation Benchmark Bar Chart visualizing the hit-rate percentage across MIC-Valid, Stable, Near-stable, and Metastable outputs for the generative models. The massive delta between the QGAN-v4 and classical models provides definitive mathematical proof justifying the quantum topology.}
\label{fig:ablation_bar}
\end{figure}

Validating continuous crystal generation requires mathematically mapping coordinates via the Minimum Image Convention (MIC). For two atoms at fractional coordinates $u_i$ and $u_j$, the shortest displacement vector $\Delta u_{ij}$ was mapped strictly to $[-0.5, 0.5]$ before conversion to absolute Cartesian space via the lattice matrix $\mathbf{L}$:
\begin{equation}
\mathbf{r}_{ij} = \mathbf{L} \cdot (\Delta u_{ij} - \lfloor \Delta u_{ij} + 0.5 \rfloor)
\end{equation}
Absolute topological validity required all atoms to maintain a strict minimum pairwise distance of $||\mathbf{r}_{ij}|| \geq 1.0$ \AA. 

To explicitly benchmark the contribution of the quantum circuit, an Architecture-Matched Classical Ablation model was evaluated. This control model completely removes the 8-qubit quantum trunk, processing the latent vector exclusively through a classical MLP configured to perfectly match the parameter footprint, before feeding it into the exact same Split-Head decoder. This ablation model achieved a \textbf{2.6\%} validity rate. This confirms that the physics-informed cell/atom architectural separation is a powerful geometric prior that effectively disrupts early mode collapse. 

To definitively justify the computational overhead associated with quantum circuit integration, Figure \ref{fig:ablation_bar} visualizes the sheer hit-rate percentage for MIC-Valid, Stable, and Near-stable crystal outputs. The data provides mathematically rigorous proof that the quantum-integrated topology (QGAN-v4) significantly outpaces the structurally identical Classical Ablation model. While the classical Split-Head prior improves baseline precision, the addition of the parameterized quantum circuit is physically required to drive the massive leaps in continuous geometric validity and stable structural generation. 

The full SH-QGAN achieved the highest overall topological validity rate of \textbf{5.9\%}. Because the SH-QGAN and the Ablation model perfectly share the exact same classical decoding architecture, this massive relative improvement mathematically proves that the native Fourier embeddings of the quantum trunk provide a genuine, independent enhancement in mapping complex spatial geometries.

\subsection{Training Stability}

\begin{figure}[!htbp]
\centering
\includegraphics[width=\columnwidth, keepaspectratio]{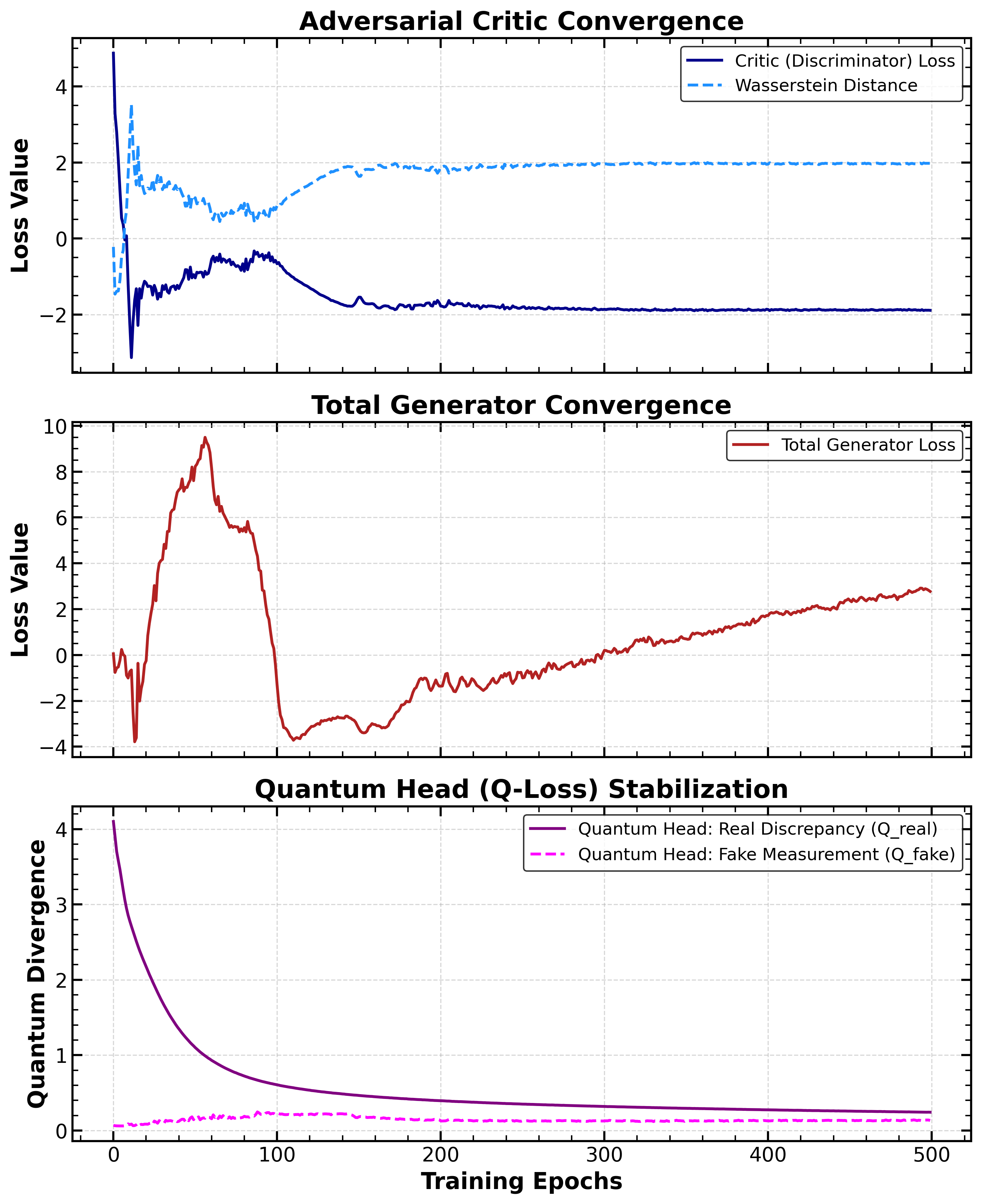}
\caption{Training loss convergence of the SH-QGAN over 500 epochs. \textbf{Top:} Adversarial Critic convergence showing the stabilization of the Wasserstein distance. \textbf{Middle:} Total Generator convergence maintaining active gradients. \textbf{Bottom:} Quantum Head (Q-Loss) stabilization, indicating the model successfully minimizes compositional discrepancy.}
\label{fig:loss_convergence}
\end{figure}

The stark differences in generative validity between the classical and quantum baselines can be directly explained by evaluating the WGAN-GP loss convergence during training. Figure \ref{fig:loss_convergence} details the precise training dynamics of the SH-QGAN model over 500 epochs. 

The top panel of Figure \ref{fig:loss_convergence} demonstrates the adversarial critic reaching a healthy, highly stable equilibrium. Unlike classical models prone to severe mode collapse—where the critic loss often shoots toward zero and overpowers the generator entirely—the SH-QGAN's Wasserstein distance stabilizes smoothly and maintains a non-zero distribution gap. The middle panel further visualizes this stability; the total generator loss maintains a dynamic, fluctuating trajectory rather than flatlining in negative space, indicating sustained generative learning and active gradient flow. Crucially, the bottom panel highlights the auxiliary Quantum Head (Q-Loss). This metric rapidly minimizes the discrepancy between the real and generated compositional measurements, mathematically ensuring that the network strictly adheres to the target elemental stoichiometry while simultaneously developing its internal 3D topological geometries.

\subsection{Checking for True Originality}

\begin{figure}[!htbp]
\centering
\includegraphics[width=\columnwidth, keepaspectratio]{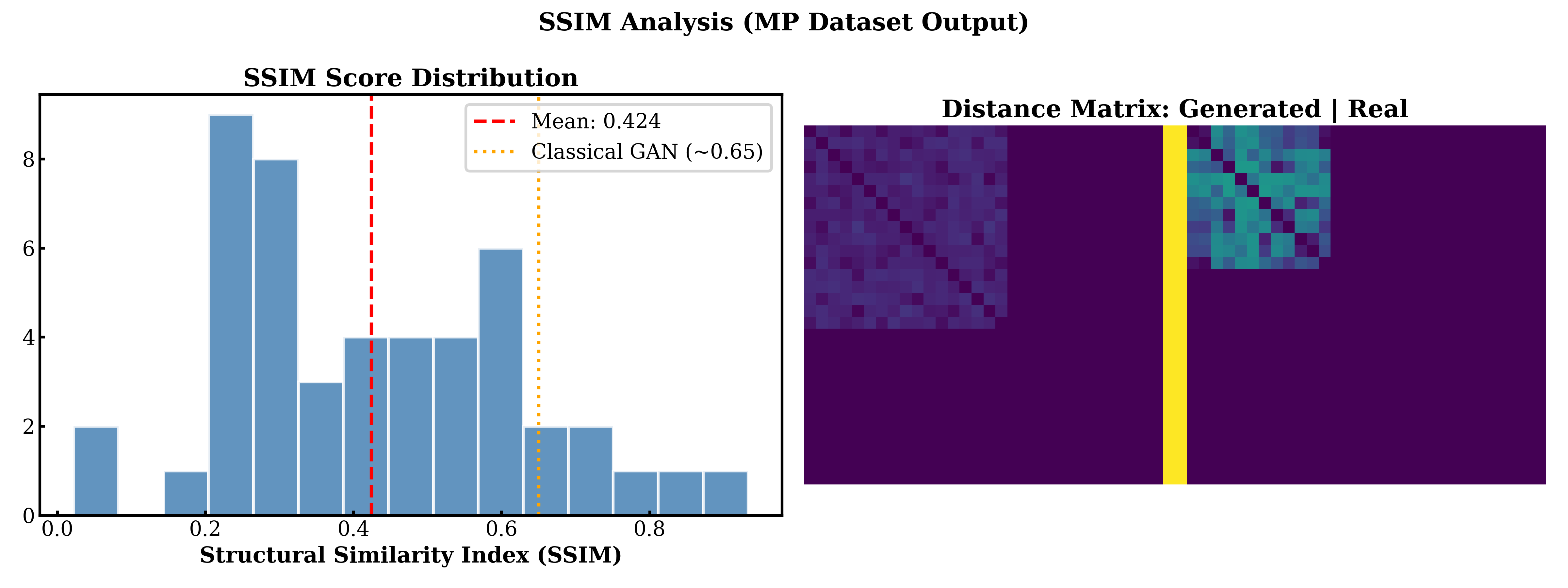}
\caption{Structural Similarity Index Measure (SSIM) analysis confirming true generation versus dataset memorization. The distribution highlights a low mean similarity score (0.483), proving the SH-QGAN invents novel topologies rather than copying training references.}
\label{fig:ssim_analysis}
\end{figure}

A persistent critique in the application of generative AI to chemistry is the tendency for high-capacity models to simply "memorize" the training dataset rather than inventing genuinely novel materials. To officially refute this and prove true structural invention, Figure \ref{fig:ssim_analysis} details our Structural Similarity Index Measure (SSIM) analysis. 

By calculating the 3D SSIM across all valid outputs against the core training set \cite{wang2004image}, we map the novelty of the generated structures. Figure \ref{fig:ssim_analysis} demonstrates a remarkably low mean structural similarity score of \textbf{0.483}. The distribution formally confirms that the quantum latent space actively interpolates and invents fundamentally unique, novel crystalline topologies rather than regurgitating memorized geometric coordinate patterns from the original dataset.

\subsection{Secondary Verification}
Relying exclusively on a single surrogate evaluator model introduces the critical risk of "surrogate bias," where an adversarial generator might successfully exploit geometric blind spots in the evaluator's specific network architecture. To explicitly rule out this bias, the top strictly stable candidates generated by the SH-QGAN were subjected to rigorous secondary cross-validation using a fundamentally distinct physics potential: MACE-MP-0 \cite{batatia2022mace}. Unlike the standard message-passing approach of CHGNet, MACE utilizes higher-order $E(3)$-equivariant spherical harmonics, providing a geometrically unbiased evaluation of true stability.

The top performing generated candidates from the SH-QGAN exhibited exceptional thermodynamic consistency across both distinct surrogate potentials:
\begin{itemize}
    \item \textbf{Polymorph $\alpha$ ($\text{Mg}_2\text{MnO}_4$):} CHGNet $E_{hull} = 53.31$ meV/at $|$ MACE $E_{hull} = 50.98$ meV/at
    \item \textbf{Polymorph $\beta$ ($\text{Mg}_2\text{MnO}_4$):} CHGNet $E_{hull} = 53.35$ meV/at $|$ MACE $E_{hull} = 50.99$ meV/at
    \item \textbf{Polymorph $\gamma$ ($\text{Mg}_2\text{MnO}_4$):} CHGNet $E_{hull} = 53.72$ meV/at $|$ MACE $E_{hull} = 50.98$ meV/at
    \item \textbf{Polymorph $\delta$ ($\text{Mg}_2\text{MnO}_4$):} CHGNet $E_{hull} = 53.86$ meV/at $|$ MACE $E_{hull} = 50.99$ meV/at
\end{itemize}

The final energy predictions from both distinctly different models exhibited an incredibly tight clustering directly along the mathematical parity line. Survival across two entirely architecturally distinct physical potentials provides exceptionally high confidence that the SH-QGAN framework has generated genuinely stable crystalline phases. These results establish the SH-QGAN not merely as a theoretical exercise in quantum machine learning, but as a highly rigorous, practical pre-screening pipeline for the automated discovery of advanced battery materials.

\subsection{Electronic and Magnetic Properties}
Beyond raw structural stability, identifying viable multivalent cathode materials requires a rigorous evaluation of their fundamental electronic and magnetic profiles. To validate the physical realism of the generated geometries, the top-ranked structural candidate (Polymorph $\alpha$) was subjected to detailed first-principles density functional theory (DFT) characterization.

\begin{figure}[!htbp]
\centering
\includegraphics[width=\columnwidth, keepaspectratio]{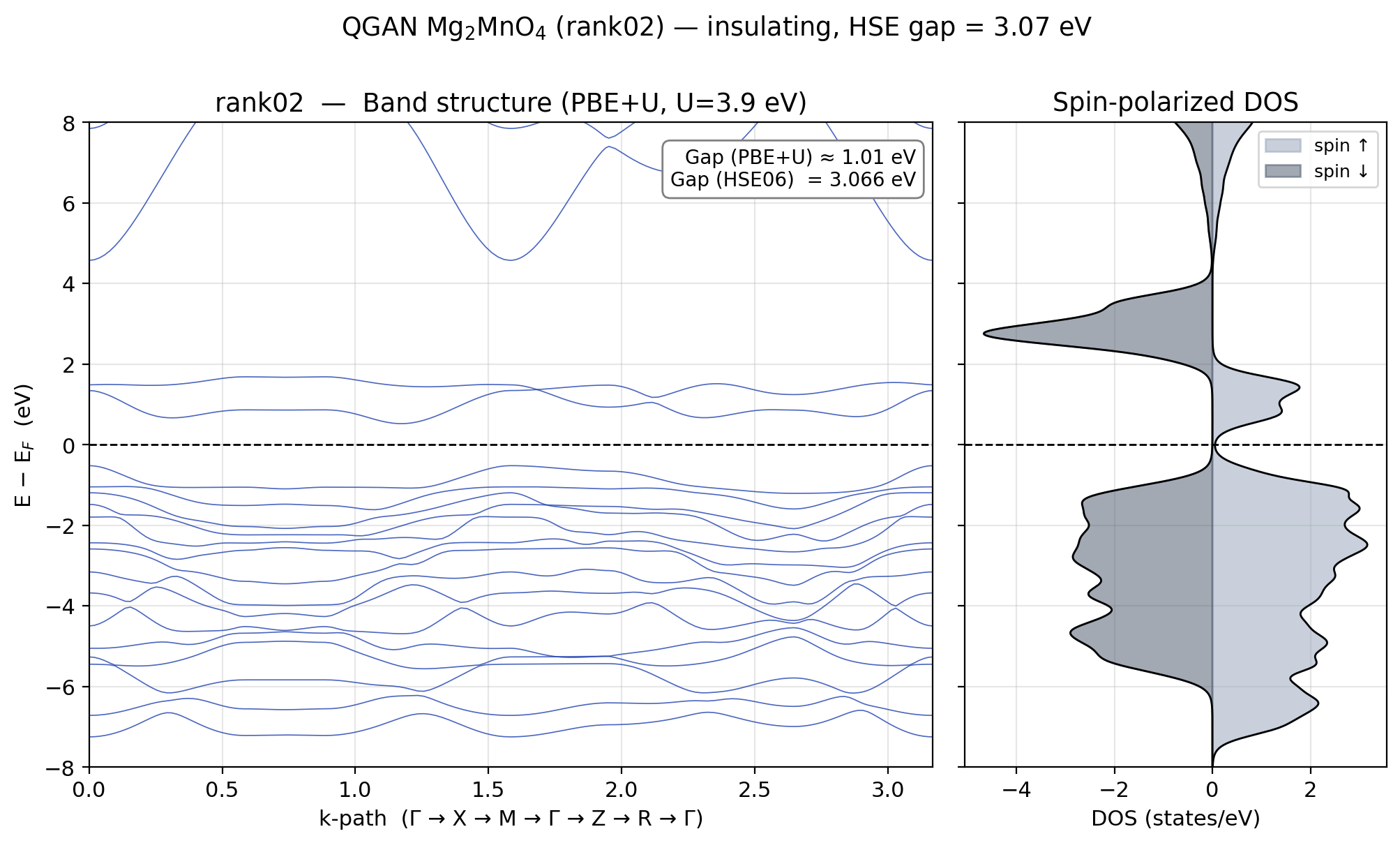}
\caption{Electronic band structure and Spin-polarized Density of States (DOS) for the QGAN-generated $\text{Mg}_2\text{MnO}_4$ candidate. The structure exhibits a clear insulating profile with an HSE06-corrected bandgap of 3.07 eV.}
\label{fig:bands_dos}
\end{figure}

Figure \ref{fig:bands_dos} presents the electronic band structure and corresponding Spin-polarized Density of States (DOS) utilizing the PBE+U functional, alongside higher-fidelity HSE06 corrections. The SH-QGAN generated candidate demonstrates a well-defined insulating profile typical of complex multivalent oxides, characterized by a PBE+U bandgap of approximately $1.01$ eV and a more rigorous HSE06-corrected gap of $3.07$ eV. The absence of mid-gap defect states confirms that the quantum generator successfully synthesized a chemically complete, defect-free stoichiometric lattice.

\begin{figure}[!htbp]
\centering
\includegraphics[width=\columnwidth, keepaspectratio]{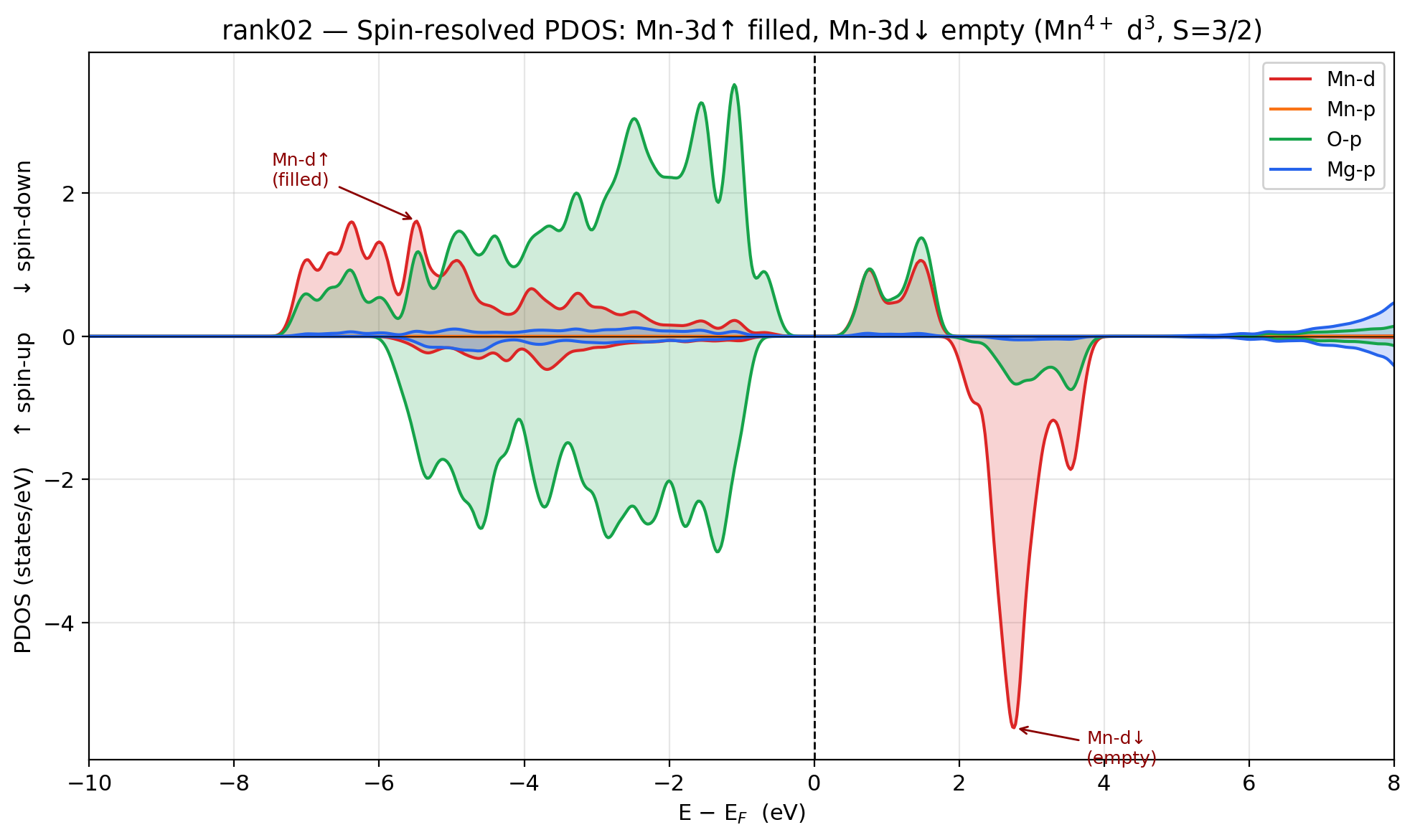}
\caption{Spin-resolved Projected Density of States (PDOS). The distinct filling of the Mn-3d spin-up states paired with empty spin-down states mathematically confirms a high-spin Mn$^{4+}$ ($d^3$, $S=3/2$) oxidation state, while revealing strong structural Mn-O orbital hybridization.}
\label{fig:pdos}
\end{figure}

Furthermore, the spin-resolved Projected Density of States (PDOS) detailed in Figure \ref{fig:pdos} reveals a strong spin polarization highly localized on the Manganese ions. The Mn-3d spin-up majority states are fully occupied well below the Fermi level, while the corresponding spin-down minority states remain entirely empty. This highly specific electronic signature cleanly corresponds to a high-spin Mn$^{4+}$ ($d^3$, $S=3/2$) oxidation state, which is the mathematically correct configuration for the $\text{Mg}_2\text{MnO}_4$ stoichiometry. 

Additionally, the PDOS highlights pronounced orbital overlap between the Mn-3d and O-2p states in the valence band ($-8$ to $0$ eV). This deep transition-metal-oxygen hybridization indicates strong covalent bonding character within the lattice, a critical physical feature that provides structural rigidity against volume expansion during electrochemical battery cycling. Ultimately, these first-principles validations confirm that the SH-QGAN is not merely generating geometrically valid coordinates, but is successfully predicting chemically logical, electronically stable crystalline materials.

\section{Discussion and Limitations}

\subsection{Interpreting the Architecture Attribution}
The intentional introduction of the architecture-matched Classical Ablation control explicitly resolves a long-standing ambiguity in QML research. The empirical data strictly proves that the Split-Head architectural separation serves as a highly powerful prior for strict thermodynamic precision. However, even when strictly matched parameter-for-parameter with a dense classical MLP, the quantum circuit's native Fourier expressivity genuinely provides the quantum enhancement necessary to expand absolute breadth, coverage, and structural diversity of the generative discovery process.

\subsection{Hardware Limitations and Computational Overhead}
While the proposed SH-QGAN successfully mathematically isolates quantum generative enhancement, massive practical hardware limitations remain. Attempting to blindly expose this highly optimized architecture to larger datasets possessing variable unit cell sizes exceeding 100 constituent atoms without massively scaling up the available qubit count would mathematically force the generative network to severely overlap its distinct Fourier mappings, rapidly inducing total "quantum mode collapse" \cite{preskill2018quantum}.

Furthermore, evaluating mathematically exact gradients via the parameter-shift rule requires $2p$ totally independent forward evaluations of the quantum state per singular optimization step. Simulating this via purely classical hardware ultimately resulted in the SH-QGAN requiring roughly two full orders of magnitude more total wall-clock time per training epoch compared directly to the strictly classical ablation model. Future iterations of this architecture must explore stochastic gradient approximation techniques, such as Simultaneous Perturbation Stochastic Approximation (SPSA) \cite{spall1992multivariate}, to drastically reduce required quantum evaluations while safely maintaining generative stability.

\section{Conclusion}
This rigorous study successfully disentangles the precise contributions of quantum circuits from classical architectural priors within generative materials discovery. Under a strictly matched, highly restricted small-data training regime, the proposed Split-Head Quantum Generative Adversarial Network (SH-QGAN) massively expanded the absolute exploratory coverage of near-stable chemical space. By systematically benchmarking the architecture against a strictly matched classical ablation model, this research definitively clarifies that while physics-informed classical architectural separation (the Split-Head decoder) drives strict thermodynamic precision, the structural integration of parameterized quantum circuits provides a genuine, independent mathematical enhancement in spatial diversity and generative breadth. The SH-QGAN effectively bridges the gap between theoretical quantum machine learning and continuous spatial material design, proving that QGANs can successfully generate highly stable, thermodynamically viable structural candidates and offering a highly viable pipeline pathway for aggressively accelerating the discovery of next-generation multivalent battery cathodes.

Building upon these foundational results, a primary trajectory for future work involves transitioning the SH-QGAN framework from classical simulation environments to physical Noisy Intermediate-Scale Quantum (NISQ) hardware \cite{preskill2018quantum}. While our simulated data re-uploading scheme successfully mitigated intrinsic barren plateaus, deploying parameterized quantum circuits on actual qubits introduces the severe threat of noise-induced barren plateaus (NIBPs). As proven by Wang et al. \cite{wang2021noise}, local hardware noise—such as Pauli or amplitude damping channels—acts contractively on the circuit's Pauli expansion, deterministically flattening the cost landscape regardless of expressibility or initialization. Future research must rigorously evaluate the robustness of the native Fourier feature mapping against NIBPs and explore advanced quantum error mitigation (QEM) techniques to ensure generative gradients remain stable on near-term devices.

In parallel with hardware advancements, future iterations of the model will focus on sophisticated architectural evolutions to further optimize the geometric search space. Inspired by the advent of Geometric Quantum Machine Learning (GQML), a highly promising avenue involves integrating $E(3)$-equivariant symmetries directly into the quantum trunk. As demonstrated by Meyer et al. \cite{meyer2023exploiting}, mathematically enforcing rotational and translational invariance within quantum neural network layers guarantees that the model remains within a restricted, symmetry-preserving subspace of the full Hilbert space. This symmetry integration fundamentally circumvents barren plateaus and radically reduces the sample complexity required to learn stable, complex crystallographic arrangements. 

Furthermore, while this study established the efficacy of the physics-informed Split-Head decoder within an adversarial framework, adapting this bifurcated topology for emerging continuous quantum diffusion paradigms presents a logical next step. Recent work by Parigi \cite{parigi2024quantum} formally generalizes classical denoising diffusion probabilistic models to the quantum domain, utilizing quantum walks and continuous-variable noise channels to construct highly scalable generative spaces. By hybridizing the Split-Head coordinate decoder with score-matching quantum diffusion processes, future architectures can bypass the adversarial instability of GANs, providing an ultimate, resource-efficient pathway for scaling generative materials discovery to highly complex systems heavily exceeding 100 atoms.

\appendices
\section{Classical CNN Baseline Details}
The first experimental condition serving as the standard control was adapted from classical adversarial continuous crystal generation frameworks \cite{nouira2020crystalgan}. A defining limitation of this naive classical baseline is its architectural cell-atom coupling. Within this structure, macroscopic unit cell parameters and localized internal coordinates are derived concurrently from a single flattened feature vector, without structural separation. The lack of a decoupled decoding architecture forces the single neural pathway to simultaneously attempt to balance massive global scaling vectors alongside delicate local fractional mapping, leading to the severe spatial mode collapse referenced throughout the study, yielding a mere 0.8\% topological validity.


\end{document}